\documentclass[conference]{IEEEtran}
\usepackage{multirow}
\usepackage[dvips]{graphicx}
\usepackage{amsmath}
\usepackage[psamsfonts]{amssymb}
\usepackage{amsxtra}
\usepackage{graphicx,color}
\usepackage{threeparttable}
\usepackage{url}
\usepackage{CJKutf8}
\usepackage{epsfig,epstopdf}
\usepackage[hang]{footmisc}

\AtBeginDvi{\input{zhwinfonts}}

\usepackage{fancyhdr}

\begin{document}
\title{Local Training for PLDA in Speaker Verification}

\author{%
\IEEEauthorblockN{%
Chenghui Zhao\IEEEauthorrefmark{1},
Lantian Li\IEEEauthorrefmark{2},
Dong Wang\IEEEauthorrefmark{2},
and April Pu\IEEEauthorrefmark{1}
}

\IEEEauthorblockA{%
\IEEEauthorrefmark{1}
Pachira, Beijing, China \\
E-mail: {\{zhaochenghui, april\_pu\}@pachiratech.com}
}

\IEEEauthorblockA{%
\IEEEauthorrefmark{2}
Center for Speech and Language Technologies, Division of Technical Innovation and Development, \\
Tsinghua National Laboratory for Information Science and Technology;\\
Center for Speech and Language Technologies, Research Institute of Information Technology;\\
Department of Computer Science and Technology, Tsinghua University, Beijing, China\\
E-mail: {lilt@cslt.riit.tsinghua.edu.cn}
}
Corresponding Author: {wangdong99@mails.tsinghua.edu.cn}
}

\maketitle

\begin{abstract}
  PLDA is a popular normalization approach for the i-vector model,
  and it has delivered state-of-the-art performance in speaker verification.
  However, PLDA training requires a large amount of labeled development
  data, which is highly expensive in most cases.
  A possible approach to mitigate the problem is various unsupervised
  adaptation methods, which use unlabeled data to adapt the PLDA scattering matrices
  to the target domain.

  In this paper, we present a new `local training' approach that
  utilizes inaccurate but much cheaper local labels to train the PLDA model.
  These local labels discriminate speakers within a single conversion only, and
  so are much easier to obtain compared to the normal `global labels'.
  Our experiments show that the proposed approach can deliver significant
  performance improvement, particularly with limited globally-labeled data.

\end{abstract}



\IEEEpeerreviewmaketitle

\section{Introduction}

The i-vector model plus various normalization approaches offers the standard framework
for modern speaker verification systems~\cite{dehak2011front,garcia2011analysis,lei2014novel,kenny2010bayesian}.
Basically, the i-vector model uses a Gaussian mixture model (GMM) or a deep neural network (DNN)
to collect the Baum-Welch sufficient statistics of an utterance, and then projects it onto a
low-dimensional total variability space. These low-dimensional representations, or i-vectors,
involve mixed information from both speakers and channels,
and therefore require some normalization techniques to separate speaker information
from other undesired variability.
Probabilistic linear discriminant analysis (PLDA) is one of the most popular normalization methods.
It assumes that i-vectors of utterances of a particular speaker form a Gaussian
distribution, with the mean vector following a normal distribution~\cite{garcia2011analysis}.
Scoring the hypothesis that two i-vectors belong to the same speaker with the PLDA model involves
marginalization of the prior distribution under the two hypothesises that the two i-vectors
are from the same speaker or not.
Combined with length normalization, PLDA delivers state-of-the-art performance in
various test benchmarks~\cite{kenny2010bayesian}.

Training a PLDA model requires a large amount of labeled data, usually thousands of speakers,
each with multiple sessions.
For example, in the two popular development databases Fisher~\cite{fisher2004} and Switchboard~\cite{switch},
there are 12,399 and 543 speakers, respectively.
In many practical situations, collecting such a large amount of labeled data is very difficult and time consuming.
For instance, for a phone-call archive from a call-center service, it is often highly difficult and time consuming to tell
whether two calls are from the same speaker, and it is more difficult to cluster calls from customers
into thousands of speakers.
Therefore, a typical situation that we often encounter is: a small labeled database is available, but it is often out-of-domain;
on the other hand, there is a large amount of in-domain data but the data are difficult to label.
This situation is also reflected in the NIST i-vector challenge~\cite{language}, where 36,572 unlabeled
i-vectors were provided for system building. How to use unlabeled data is a critical problem in
particular for practical systems.

A number of techniques have been proposed to deal with the situation, most of them are based on unsupervised adaptation.
For example, Garcia-Romero et al.~\cite{garcia2014improving} used an out-of-domain PLDA to cluster in-domain
data into some pseudo speakers, based on which the PLDA covariance matrices were adapted.
Villalba and colleagues~\cite{villalba2014unsupervised} proposed a variational Bayesian method
where the unknown labels were treated as latent variables.
Liu et al.~\cite{liu2014utilization} proposed an approach where unlabeled data (i-vectors) were treated
as from a universal speaker. The i-vectors of the universal speaker and other speakers were pooled
together to train the PLDA model.
Wang et al.~\cite{wang2016domain} proposed a domain-adaptation approach based on maximum likelihood
linear transformation (MLLT), and Rahman et al.~\cite{rahman2015dataset} proposed a dataset-invariant
covariance normalization approach that normalized i-vectors by a global covariance matrix
computed from both in-domain and out-domain data. This is equal to project i-vectors of in-domain
and out-domain speakers onto a third dataset-invariant space, so the PLDA model trained with
the projected i-vectors is more robust against data mismatch.

In this paper, we propose a new PLDA training approach that is different from the above methods.
The basic idea is to use the prior knowledge that a conversation involves only a few participants,
and these participants can be easily separated by listeners or any audio segment method, resulting
in conversation-based labels.
These labels, however, only valid within individual conversions as they do not consider anything
about the labels in other conversations. To obtain labels that can be used for PLDA training,
we further assume that all speakers in any two conversations are distinct, leading to speaker
labels something like `\emph{conversation-id:speaker-id-in-conversation}'.
This assumption, of course, is not certainly true, because it is very possible that one speaker
appears in multiple conversations. However, we find that in some practical scenarios, the possibility
that the same speaker appears in two or more conversions is rather low. For example, in a call
center archive of one week, customers in different conversations are almost different.
We call the speaker labels that only discriminates participants within individual conversions
as \emph{local labels}, and the conventional labels that accurately discriminate cross-conversation
speakers as \emph{global labels}. The PLDA training with local labels is identical
to the procedure with global labels.
Note that a major difficulty for speaker labeling
is the comparison for speakers within different conversations, which means local labels are much
cheaper than global labels, although they are not thus accurate and can be only regarded as
partial supervision.

This paper is structured as follows: Section~\ref{sec:local} presents the local training approach,
and Section~\ref{sec:exp} presents the experiments, followed by some conclusions
in Section~\ref{sec:conc}.

\section{Local PLDA training}
\label{sec:local}

In this section, the conventional PLDA model is briefly reviewed, and our proposed local PLDA
training approach is then presented in details.

\subsection{PLDA model}

PLDA is an extension of the linear discriminative analysis (LDA), by introducing a Gaussian
prior on the mean vector of classes. Combined with length normalization, PLDA has
delivered state-of-the-art performance in speaker verification.
Letting $w_{ij}$ denote the i-vector of the $j^{th}$ utterance (session) of the $i^{th}$ speaker,
the PLDA model can be formulated as follows:

\vspace{-2mm}
\[
w_{ij}  =  u + V y_{i} + z_{ij}
\]

\noindent where $u$ is the speaker-independent global factor, and $y_{i}$ and $z_{ij}$ represent the speaker-level
and utterance-level factors, respectively. The matrix $V$ involves the bases of the speaker subspace. Note that
both $y_i$ and $z_{ij}$ are assumed to follow a full-rank Gaussian prior.
The model can be trained via an EM algorithm~\cite{prince2007probabilistic},
and the similarity of two i-vectors can be computed as the ratio of the evidence (likelihood) of
two hypothesises: whether or not the two i-vectors belong to the same speaker~\cite{Yang2012plda}.

\subsection{Global training and local training}

\begin{figure*}[htop]
\begin{center}
\includegraphics[width=0.9\linewidth]{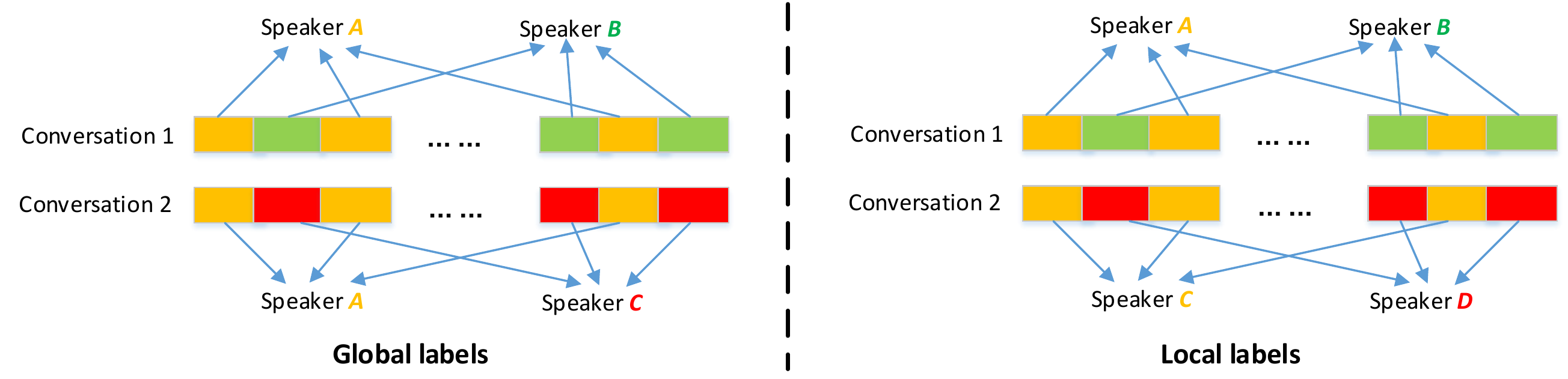}
\end{center}
\caption{Illustration of the difference between local labels and global labels.}
\label{fig:labels}
\end{figure*}

Training a PLDA model also requires a large amount of labeled data,
usually thousands of speakers each with multiple sessions.
These labels discriminate speakers within multiple sessions and so are
global labels. Global labels are very expensive, because it is usually
very hard to tell whether a voice from a new utterance is from a speaker
in a set that involves thousands of speakers that are already known, or from
a new speaker. Most of existing databases were collected following
a registration-and-recording approach, which identifies speaker identities
by meta information, instead of manual labeling. This approach is cheap
in data labeling, but is costly in hiring speakers and managing the recording.
Furthermore, it is not applicable in many practical scenarios where
some in-domain data are important and therefore should be collected in
a real-life environment but the meta information is not available.

Some unsupervised learning methods have been proposed to solve the problem,
as discussed already in the introduction~\cite{garcia2014improving,villalba2014unsupervised,liu2014utilization,wang2016domain,rahman2015dataset}.
Basically, these methods
focus on i-vector normalization or adaptation, so that the
normalized or adapted i-vectors can be better discriminated by the PLDA model
that has been trained already. In other words, they can not improve the discriminative
capability of the PLDA model.

We propose a local training approach that can be regarded as a weak-supervised
method. Basically we label speakers in a conversation-independent way, which means that
the labels only discriminate speakers within the same conversion, and speakers
in different conversations are simply assumed to be different. With the local
labels obtained, PLDA is trained as usual. We call this training based on local
labels \emph{local training}.
Although this supervision is not fully accurate, we hope it is still
possible to improve PLDA.

Fig.~\ref{fig:labels} illustrates the difference between local labels and
global labels, where each speaker is represented by a particular color.
For global labels, the segments from the same speaker but different conversations
are correctly labeled. For local labels, speakers in different conversations
are labeled as distinct.

\section{Experiment}
\label{sec:exp}

The proposed local training approach is tested on a speaker verification task
with telephone speech from a call-center archives. The system is designed
based on the GMM-ivector framework. We first present the data profile and then
report the results. Some analysis will be given to show in which
condition the local training is most effective.

\subsection{Databases and experimental setting}

The training data used to train the GMM-ivector system are composed of $500$ hours
of conversational speech signals. These data are used to train the
UBMs and the T matrix of the i-vector model.
The development data used to train the PLDA model are divided into two sets:
the Global set and the Local set, with global and local labels respectively.
Note that, the environmental condition of the Local set is more close
to the condition of the evaluation data, which means that the Local set can be
regarded as in-domain data and the unsupervised learning would be helpful.
More details about the development data are shown in Table~\ref{tab:dev}.

\begin{table}[htb]
\normalsize
\centering
\caption{Development set for PLDA training.}
\label{tab:dev}
\begin{tabular}{l|c|c}
\hline
                        &    $\#$ of Spks &  $\#$ of Utts     \\
\hline
\hline
    \textbf{Global}     & 6,000         &    42,719           \\
\hline
    \textbf{Local}      & 5,532         &    64,943           \\
\hline

\end{tabular}
\end{table}

The evaluation set involves $1,236$ speakers and the enrollment speech for each speaker
is $15$ seconds long. The test is conducted in $3$ conditions, where the length of
the test utterances grows from $5$ seconds to $15$ seconds. The details of the
evaluation data are shown in Table~\ref{tab:cond}.

\begin{table}[htb]
\normalsize
\centering
\caption{Evaluation set}
\label{tab:cond}
\begin{tabular}{l|c|c|c|c}
\hline
                 & $\#$ of Spks &  $\#$ of Utts   & Duration(s) &  $\#$ of Trials \\
\hline
\hline
   \textbf{C5}   & 1,236        & 3,708            &      5       &   4,583,088   \\
\hline
   \textbf{C10}   & 1,236        & 2,472            &     10       &   3,055,392 \\
\hline
   \textbf{C15}   & 1,236        & 2,472            &     15       &   3,055,392 \\
\hline

\end{tabular}
\end{table}

The acoustic feature used in our experiments is the $60$-dimensional Mel frequency cepstral coefficients (MFCCs),
which involves $20$-dimensional static components plus the first and second order derivatives.
The frame size is $25$ ms and the frame shift is $10$ ms. The UBM
involves $1,024$ Gaussian components and the dimensionality of the i-vectors is $100$.
The performance is evaluated in terms of Equal Error Rate (EER)~\cite{greenberg20132012}.

\subsection{Basic results}
\label{sec:basic}

We test three training strategies for PLDA, as shown below:

\begin{itemize}
\item \textbf{Global training (GT)}: The conventional PLDA training with the Global dev set.
It is regarded as the baseline in our experiment.
\item \textbf{Local training  (LT)}: Local PLDA training with the Local dev set.
\item \textbf{Pooled training (Pool)}: PLDA training with both the Global set and the Local set.
\end{itemize}

\begin{table}[htb]
\normalsize
\begin{center}
\caption{EER(\%) results of various recognition systems}
\label{tab:eer}
\begin{tabular}{|l|c|c|c|}
\hline
             & \multicolumn{3}{|c|}{EER\%} \\
\hline
             &   C5   &    C10   &    C15    \\
\hline
Cosine       &  4.72  &   2.91  &   2.67   \\
\hline
GT           &  \textbf{2.56}  &   \textbf{1.86}  &   \textbf{1.82}    \\
\hline
LT           &  3.67  &   2.47  &   2.27   \\
\hline
Pool         &  2.72  &   1.90  &   1.90   \\
\hline
\end{tabular}
\end{center}
\end{table}

The results are shown in Table~\ref{tab:eer}. For comparison, the results with
cosine scoring are also reported. We first observe that all the three PLDA
training approaches obtained significant performance improvement compared to the
cosine scoring. This is particular interesting for the LT approach, where
only local labels are available. This confirms our conjecture that cheap local labels
can be used to train PLDA and obtain performance improvement with
little effort on data labeling.

At the same time, it can be observed that the global training (GT) is still
the most effective, and the local training (LT) and the pool training (Pool)
are unable to beat the GT. This should be attributed to the noise in local labels,
caused by the fact that the same speakers appeared in different conversations
are simply labeled as distinct speakers.


\subsection{Study on pooled training}

The superior performance with GT over LT is expected, due to the more accurate supervision with
global labels. However, the lower performance with the pooled training compared to the GT is
a bit surprising. As we have argued, the supervision with local labels is noisy but informative,
which can be seen from the LT results in Table~\ref{tab:eer}. One reason that the performance was
deteriorated is that the global training is so strong ($6000$ speakers in the Global set) that the
noisy local training is not necessary. More investigations are required to confirm the conjecture and
experiment with the condition under which local training is effective.

\begin{figure}[htop]
\begin{center}
\includegraphics[width=0.9\linewidth]{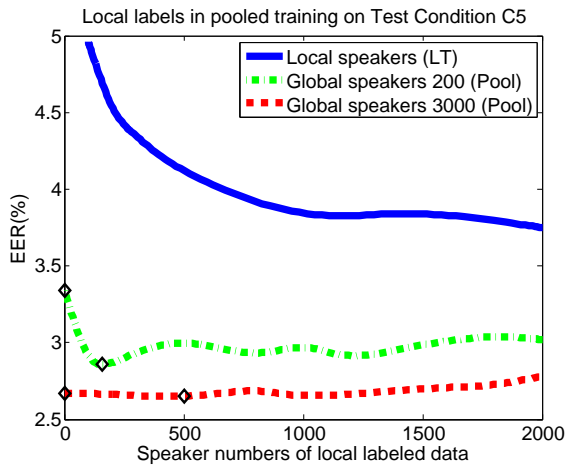}
\end{center}
\caption{Local labels in pooled training on Test Condition C5.}
\label{fig:local}
\end{figure}

We first investigate the performance change with different amount of locally labeled data, with
the globally labeled data fixed. The results are shown in Fig.~\ref{fig:local}. For a clear presentation,
we only show the results on the test condition C5; the performance on C10 and C15 show similar trends.
In Fig.~\ref{fig:local}, the number of speakers of the globally labeled data (global speaker) is set
to $0$, $200$ and $3000$ respectively, corresponding to the three curves in the picture.
The number of speakers of local labeled data (local speaker) varies from $0$ to $2000$.
Note that the case of $0$ global speaker is just the local training
approach, and the case of $0$ local speaker is simply the global training
approach.

From the results shown in Fig.~\ref{fig:local}, we first observe that the performance
of the local training approach is monotonically improved with more locally labeled data.
When the locally labeled data is sufficient, the performance of the pooled training
seems saturated at a level close to the performance of global training with hundreds of
global speakers.
Moreover, it can be seen that when the number of global speakers is $200$ (the dash line),
involving locally labeled data is helpful. The performance is firstly improved when a
small number of local speakers are involved, and then it is degraded a little when more
local speakers are involved. This result indicates that the information conveyed by
locally labeled data is useful when the globally labeled data are insufficient.
When the number of global speakers is $3000$ (the dot line), involving locally
labeled data does not improve the performance; actually it may deteriorate
the performance if the locally labeled data are too many.

\begin{figure}[htop]
\begin{center}
\includegraphics[width=0.9\linewidth]{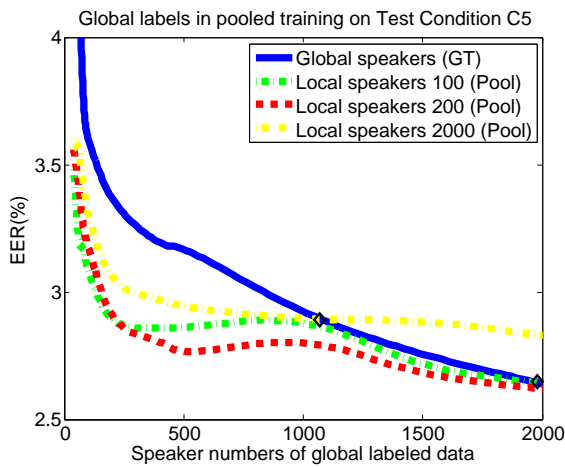}
\end{center}
\caption{Global labels in pooled training on Test Condition C5.}
\label{fig:global}
\end{figure}

To make the picture complete, we fix the number local speakers, and vary the number
of global speakers from $0$ to $2000$. The results are shown in Fig.~\ref{fig:global}.
Again, only the results on the C5 test condition are presented. The four curves
in Fig.~\ref{fig:global} show the results with the number of local speakers set to $0$,
$100$, $200$ and $2000$, respectively. The same information can be read as from Fig.~\ref{fig:local}.

As a summary, we find that the local training is mostly effective when the global training
is weak, i.e., the number of global speakers is small. If the globally labeled
data are sufficient, the local training is not very useful.
In practice, it is often the case that globally labeled data are very limited,
suggesting the potential value of the local training approach.

\section{Conclusion}
\label{sec:conc}

This paper proposed a local training approach for PLDA and verified its potential in speaker verification.
Based on the assumption that speakers in different conversations tend to be distinct,
local labels posses a high probability to be correct and so can be used as weak supervision to
train PLDA.
Experimental results demonstrated that the local training approach can improve
system performance when globally labeled data are limit.
A particular problem of the local training is the conversion independent assumption. Future work
will investigate to what extent this assumption holds would result in an effective local training.

\section*{Acknowledgment}

This work was supported by the National Natural Science Foundation of China under Grant No.61271389 and NO.61371136 and the National Basic Research Program (973 Program) of China under Grant No.2013CB329302.

\bibliographystyle{IEEEtran}
\bibliography{local}

\end{document}